\documentclass[3p,times,sort&compress]{elsarticle}
\usepackage{graphicx}
\usepackage{epsfig,amsmath,amssymb,graphicx,lscape,color}

\newcommand{\bm}[1]{\mbox{\boldmath $#1$}}
\newcommand{\grad}{\bm{\nabla}}

\journal{Physica D: Nonlinear Phenomena.}
\begin{document}

\begin{frontmatter}
\title{Hole-closing model reveals exponents for nonlinear degenerate diffusivity functions in cell biology}

\author[qut]{Scott W McCue \corref{cor1}}
\author[qut]{Wang Jin}
\author[qut]{Timothy J Moroney}
\author[ntu]{Kai-Yin Lo}
\author[ntu]{Shih-En Chou}
\author[qut]{Matthew J Simpson}
\address[qut]{School of Mathematical Sciences, Queensland University of Technology, Brisbane QLD 4000, Australia.}
\address[ntu]{Department of Agricultural Chemistry, National Taiwan University, Taipei 10617, Taiwan}
\cortext[cor1]{Corresponding author: scott.mccue@qut.edu.au}
\cortext[cor2]{SWM and WJ are joint first authors.}

\begin{abstract}
Continuum mathematical models for collective cell motion normally involve reaction-diffusion equations, such as the Fisher-KPP equation, with a linear diffusion term to describe cell motility and a logistic term to describe cell proliferation.
While the Fisher-KPP equation and its generalisations are commonplace, a significant drawback for this family of models is that they are not able to capture the moving fronts that arise in cell invasion applications such as wound healing and tumour growth.  An alternative, less common, approach is to include nonlinear degenerate diffusion in the models, such as in the Porous-Fisher equation, since solutions to the corresponding equations have compact support and therefore explicitly allow for moving fronts.
We consider here a hole-closing problem for the Porous-Fisher equation whereby there is initially a simply connected region (the {\em hole}) with a nonzero population outside of the hole and a zero population inside.
We outline how self-similar solutions (of the second kind) describe both circular and non-circular fronts in the hole-closing limit.
Further, we present new experimental and theoretical evidence to support the use of nonlinear degenerate diffusion in models for collective cell motion.
Our methodology involves setting up a two-dimensional wound healing assay that has the geometry of a hole-closing problem, with cells initially seeded outside of a hole that closes as cells migrate and proliferate.
For a particular class of fibroblast cells, the aspect ratio of an initially rectangular wound increases in time, so the wound becomes longer and thinner as it closes; our theoretical analysis shows that this behaviour is consistent with nonlinear degenerate diffusion but is not able to be captured with commonly used linear diffusion.
This work is important because it provides a clear test for degenerate diffusion over linear diffusion in cell lines, whereas standard one-dimensional experiments are unfortunately not capable of distinguishing between the two approaches.

\end{abstract}

\end{frontmatter}

\section{Introduction}

The use of reaction-diffusion equations in mathematical biology is widespread~\cite{murray01,britton86}, especially for models of collective cell motion in applications of wound healing \cite{tranquillomurrary92,olsenetal95,murphy12} and tumour invasion \cite{gatenbygawlinski96,rooseetal07,sherratt01}.  The most commonly used type of diffusion in these models is Fick's first law, which gives rise to a linear diffusion term.  However, a deficiency in this approach is that such governing equations do not allow for explicit descriptions of invading fronts.  In an attempt to address this deficiency, some literature covering mathematical models of cell migration and proliferation has included studies of reaction-diffusion equations with nonlinear degenerate diffusion \cite{bakersimpson12,jinetal16,mainietal04a,sherrattmurray90a,sengersetal07,simpsonetal2011,warne2019,simpsonetal2010a}.  These models allow for solutions with a well-defined moving boundary at the front of the invading cell population.  An ongoing challenge in this area of research is to identify the most appropriate choice of nonlinear diffusion.  We continue this work in the present paper by applying analytical and numerical techniques to study a hole-closing problem in the plane, focussing on the role of nonlinear degenerate diffusion and supporting our study with experimental results from an {\em in vitro} cell migration assay (that complements our recent study~\cite{jinetal18}).

In this paper we study the most simple nontrivial reaction diffusion equation with nonlinear degenerate diffusion that describes cell proliferation and migration, namely the Porous-Fisher equation
\begin{equation}
\frac{\partial u}{\partial t}=D\,\grad\cdot\left(\left(\frac{u}{K}\right)^n\grad u\right)+\lambda u\left(1-\frac{u}{K}\right).
\label{eq:PFE}
\end{equation}
This is a parabolic partial differential equation with nonlinear degenerate diffusion and a logistic growth term~\cite{atkinsonetal81,dePablo98,harris04,hilhorstetal08,kingmccabe03,medvedev03,newman80,sanchezgarduno95,sherrattmarchant96,witelski95}.  In cell biology, $u$ represents the density of a particular cell type, while ${\bf x}\in\mathbb{R}^N$, where $N=1$, 2 or 3.  The diffusion term in (\ref{eq:PFE}) describes migration (or cell motility) of the cell population, with $n>0$ corresponding to a scenario in which cells are more likely to migrate if crowded, while the logistic growth component of (\ref{eq:PFE}) models cell proliferation with a carrying capacity $K$.  The limiting case $n=0$ reduces (\ref{eq:PFE}) to the well-known the Fisher-Kolmogorov-Petrovski-Piskunov (Fisher-KPP) equation
\begin{equation}
\frac{\partial u}{\partial t}=D\,\nabla^2 u+\lambda u\left(1-\frac{u}{K}\right),
\label{eq:FE}
\end{equation}
for which cell migration is due to random cell motility that is independent of cell density.  As mentioned above, there are a plethora of studies in cell biology and ecology that use (\ref{eq:FE}) or related models.  In much of this vast literature, the Fisher-KPP equation has proved effective, and can successfully reproduce experimental behaviours; however, the choice of linear Fickian diffusion (or $n=0$ in (\ref{eq:PFE})) is often also made because of its simplicity, not necessarily because of its biological relevance.

The main goal of the present study is to explore the role of nonlinear degenerate diffusion in a simple model of collective cell motion.  As such, it is important to emphasise the key feature of the Porous-Fisher equation (\ref{eq:PFE}) with $n>0$ is that, unlike the Fisher-KPP equation (\ref{eq:FE}) with linear diffusion, it allows solutions with compact support.  This is a direct consequence of the term $(u/K)^n$, which vanishes in the limit $u\rightarrow 0^+$ (the diffusion is said to be {\em degenerate} in the sense that as $u\rightarrow 0^+$, the diffusion itself vanishes, and the equation changes from being of parabolic type to elliptic; to see this, write $v=u^n$ to give the eikonal equation $\partial v/\partial t\sim |\grad v|^2$ in the limit $u\rightarrow 0^+$).  Therefore, it is possible to use the Porous-Fisher equation (\ref{eq:PFE}) to model well-defined fronts of cell populations advancing on a region of zero population.  On the other hand, even with initial conditions that have compact support, solutions to the Fisher-KPP equation (\ref{eq:FE}) have $u>0$ for all ${\bf x}$ and $t>0$, meaning information is travelling infinitely fast, which strictly speaking is not biologically realistic.

\begin{figure}[p]
\centering
\includegraphics[width=0.75\textwidth]{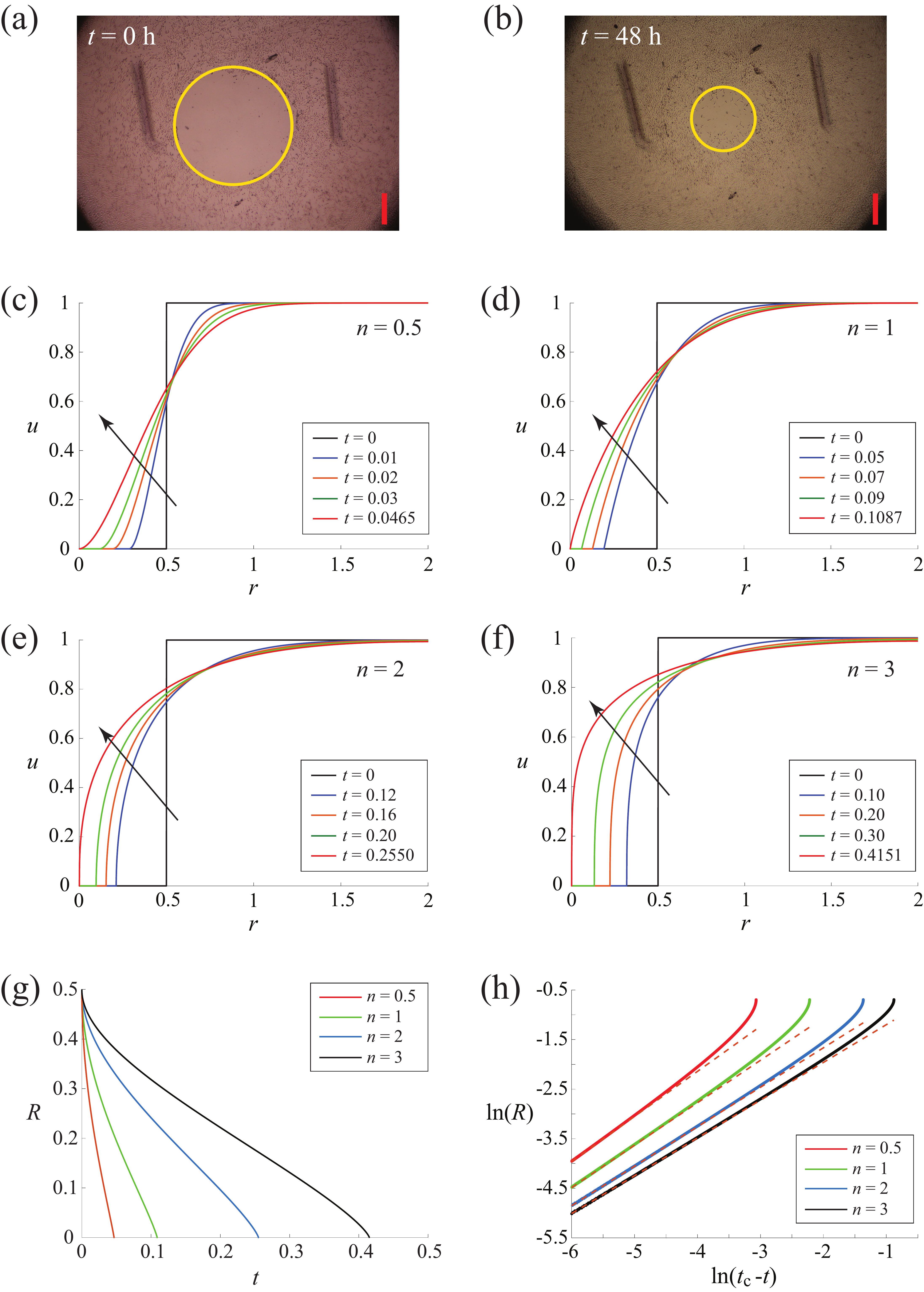}
\caption{{\bf Experimental images of hole-closing and numerical solutions.} (a)--(b) Experimental images from a sticker assay with the position of the leading edge highlighted by yellow circles. The circular holes at $t = 0$ and 48 h are shown. The scale bar corresponds to 500 $\mu$m. (c)--(f) Density profiles for the dimensionless Porous-Fisher equation with $n = 0.5, 1, 2,$ and 3, respectively. The red solid line corresponds to the density profile at focusing time $t_\text{c}$. The black arrow indicate increasing time. (g)--(h) Time evolution of the position of the leading edge for $n = 0.5, 1, 2$, and 3. Dashed lines in (h) correspond to straight lines that merge with the time evolution of the position of the leading edge. All the numerical solutions are obtained on $0 < r < 2$ with a nonuniform mesh of 4001 nodes, with an initial circular-hole at $R(0)=0.5$.  Zero net flux boundary conditions are imposed at both boundaries $r=0$ and $2$.}
\label{fig:figure1}
\end{figure}

The application we have in mind is a two-dimensional cell migration assay (${\bf x}\in\mathbb{R}^2$), with an invading population of cells moving over a substrate with a sharp front, ahead of which the cell population is essentially zero.  In particular, we focus on the geometry in which there is initially a simply connected region devoid of cells, which we refer to as a {\em hole} or a {\em wound}.  The resulting hole-closing problem is (\ref{eq:PFE}) subject to initial conditions
\begin{equation}
u({\bf x},0)=I({\bf x}),
\quad \mbox{with}\quad I({\bf x})=0
\quad \mbox{for}\quad {\bf x}\in \Omega(0),
\label{eq:ichole2D}
\end{equation}
where the simple closed curved $\partial\Omega(0)$ describes the initial shape of the hole (or wound).   The challenge is to solve for $u$, but also to track the shape and speed of the boundary of the hole $\partial\Omega(t)$, especially in the hole-closing limit $t\rightarrow t_c^-$.  We are motivated by our recent (\cite{jinetal18}) and new experimental data from a wound healing ({\em sticker}) assay, as illustrated in Figure~\ref{fig:figure1}(a)-(b).  In this particular example, the initial wound is circular in shape; however, a feature of our experimental design is that we are able to make the initial wounds any shape we choose. More generally, the study of hole-closing problems in cell biology has applications to a range of experimental wound healing scenarios, such as the circular wounds created by barrier assays~\cite{ashby,kramer,treloarsimpson14}, in a mammal's ear \cite{vandenbrenk1956} or cornea~\cite{buck1979,sheardowncheng96} or a human skin equivalent construct~\cite{xie2010}, for example; similar mechanisms and geometry are involved in cell bridging experiments in tissue engineering~\cite{totti}.  Further motivation for our study arises from the lack of consensus when using (\ref{eq:PFE}) to model cell migration on the appropriate choice of the diffusion exponent $n$; we aim to shed light on this issue.

The outline of our paper is as follows.  In the following section we set up our subsequent analysis by nondimensionaling the problem and presenting illustrative numerical solutions for the radially symmetric case of a circular hole (the numerical scheme is summarised in the Supplementary Material).  In section~\ref{sec:selfsimilar}, we argue that for times leading up to the hole-closing time $t_c$, the asymptotic behaviour of (\ref{eq:PFE})-(\ref{eq:ichole2D}) in the neighbourhood of the point at which the hole closes ${\bf x}_c$ is equivalent to that for the Porous Medium equation (which is (\ref{eq:PFE}) with $\lambda=0$).  For the radially symmetric case, we revisit the role of self-similarity of the second kind for solutions to the Porous Medium equation, reproducing some existing results by using an approach which is in some sense more transparent than that documented in the literature \cite{aronsongraveleau93,angenentaronson95a,aronsonetal03}.
Numerical solutions are presented that illustrate how solutions to (\ref{eq:PFE})-(\ref{eq:ichole2D}) with $k$-fold symmetry may or may not evolve to a circle in the hole-closing limit, depending on the diffusion exponent $n$, as per the analogous case of the Porous Medium equation~\cite{aronsonetal03,beteluetal00}.  In section~\ref{sec:2fold}, we study solutions that are not self-similar in the hole-closing limit.  In particular, we choose initial conditions for which the hole is initially rectangular and show how the hole becomes long and thin in the hole-closing limit, again following the behaviour of the Porous Medium equation~\cite{angenentetal01}.  By comparing with new experimental data from recently developed sticker assays (described in the Supplementary Material), we demonstrate how to estimate the exponent $n$ in (\ref{eq:PFE}).  Finally, we close the paper in section~\ref{sec:discussion} with a summary and a discussion about how our results provide support for the use of nonlinear degenerate diffusion in models for collective cell motion.

\section{Governing equations and preliminary numerical results}\label{sec:formulation}

\subsection{Nondimensionalisation}

As discussed in the Introduction, the model we focus on is the Porous-Fisher equation (\ref{eq:PFE}), which is the simplest model for collective cell motion in a two-dimensional assay that includes nonlinear degenerate diffusion.  The only dependent variable is the cell density and, as a consequence, this model implicitly assumes that an excess of nutrients is available so that cell migration and cell proliferation are not affected by any lack of nutrient availability \cite{simpsonetal2010b}.

We nondimensionalise (\ref{eq:PFE}) by scaling cell density with respect to carrying capacity $K$ and choosing the representative length and time scales to be $\sqrt{D/\lambda}$ and $\lambda^{-1}$, respectively.  As such, our dimensionless Porous-Fisher equation,
\begin{equation}
\frac{\partial u}{\partial t}=\grad\cdot \left(u^n\grad u\right)+ u\left(1-u\right) \quad\mbox{in}\quad {\bf x}\in\mathbb{R}^2\setminus\Omega(t),
\label{eq:PMEdimensionless}
\end{equation}
does not involve any parameters apart from the diffusion exponent $n>0$.  The dimensionless version of the initial condition (\ref{eq:ichole2D}) will, however, depend implicitly on the dimensional parameters.  For example, as we discuss shortly, for a wound that is initially circular, the density $u=u(r,t)$ is a function of $r$ and $t$, and the relevant initial condition is
\begin{equation}
u(r,0)=H(r-R(0)),
\label{eq:radialIC}
\end{equation}
where $H(r)$ is the Heaviside function and $R(0)$ is the initial radius of the wound.   In this case, the dimensionless quantity $R(0)$ is the dimensional initial radius scaled by $\sqrt{D/\lambda}$.

We may formulate our hole-closing problem as a moving boundary problem by coupling (\ref{eq:PMEdimensionless}) with the conditions
\begin{equation}
u=u^n\frac{\partial u}{\partial \nu}=0 \quad\mbox{on}\quad \partial\Omega(t),
\label{eq:consmass}
\end{equation}
where here $\partial u/\partial \nu$ is a normal derivative.  The second condition in (\ref{eq:consmass}) enforces conservation of mass at the interface.

\subsection{Numerical results for radially symmetric problem}

We now provide numerical results for radially symmetric hole closing, which is governed by
\begin{equation}
\frac{\partial u}{\partial t}= \frac{1}{r}\frac{\partial}{\partial r}\left(ru^n\frac{\partial u}{\partial r}\right)
+ u\left(1-u\right),
\quad r>R(t),
\label{eq:radialPR}
\end{equation}
\begin{equation}
u=u^n\frac{\partial u}{\partial r}=0 \quad\mbox{on}\quad r=R(t),
\label{eq:consmassradial}
\end{equation}
subject to the initial condition (\ref{eq:radialIC}).  The numerical approach we use, based on applying a straight-forward finite-difference scheme on an uneven grid, is summarised in the Supplementary Material.

In Figure~\ref{fig:figure1}(c)-(f) we present density profiles for the four exponents $n=0.5$, $1$, $2$ and $3$, each with $R(0)=0.5$.  In all four cases we see that the solutions have compact support so that $u>0$ for $r>R(t)$.  Further, we note that the radius of the hole, $R(t)$, decreases in time until $R\rightarrow 0$ as $t\rightarrow t_c^-$, where $t_c$ is the hole-closing time.  The dependence of $R(t)$ on $t$ is shown in Figure~\ref{fig:figure1}(g).  We see here that the contact line $r=R(t)$ moves more slowly as the diffusion exponent $n$ increases, which is to be expected, since, for a fixed $u$, the nonlinear diffusion term $u^n$ decreases as $n$ increases.

One interesting observation is that the slope of the density profiles at the contact line $r=R(t)$ vanishes for $n<1$, is a nonzero constant for $n=1$, and is infinite for $n>1$~\cite{newman80,warne2019}.  Indeed, a simple leading order balance near the contact line for $R(t)>0$ suggests that $u\sim B(t)\,(r-R(t))^{1/n}$, which explains these qualitative behaviours.

Another observation about the propagation of the contact line is that $R(t)$ appears to follow a power-law $R(t)\sim \mu (t_c-t)^\beta$ in the hole-closing limit $t\rightarrow t_c^-$.  This point is demonstrated in Figure~\ref{fig:figure1}(h), where the solid curves are numerical results corresponding to Figure~\ref{fig:figure1}(g).  As these curves appear to approach a line on the log-log, a power-law behaviour is anticipated.  The slope $\beta$ is dependent on $n$, as we discuss in detail in the following section.

\section{Self-similar solutions}\label{sec:selfsimilar}

\subsection{Similarity solutions for radially symmetric geometry}

In the limit the hole closes, $t\rightarrow t_c^-$, we look for a similarity solution
\begin{equation}
u\sim (t_c-t)^\alpha U(\rho),
\quad\mathrm{where}\quad \rho=\frac{r}{(t_c-t)^\beta},
\label{eq:similaritysoln}
\end{equation}
where at this stage the exponents $\alpha$ and $\beta$ are unknown.  To proceed we require the partial derivatives
\begin{equation}
\frac{\partial u}{\partial t}=(t_c-t)^{\alpha-1}\left(-\alpha U +\beta\rho \frac{\mathrm{d}U}{\mathrm{d}\rho} \right),
\label{eq:dudt}
\end{equation}
\begin{equation}
\frac{\partial u}{\partial r}=(t_c-t)^{\alpha-\beta} \frac{\mathrm{d}U}{\mathrm{d}\rho},
\end{equation}
\begin{equation}
\frac{1}{r}\frac{\partial}{\partial r}\left(ru^n\frac{\partial u}{\partial r}\right)=(t_c-t)^{\alpha(n+1)-2\beta} \frac{1}{\rho}
\frac{\mathrm{d}}{\mathrm{d}\rho}\left(\rho U^n \frac{\mathrm{d}U}{\mathrm{d}\rho}\right).
\label{eq:d2udr2}
\end{equation}
We see that $\partial u/\partial t=\mathcal{O}((t_c-t)^{\alpha-1})$, $u=\mathcal{O}((t_c-t)^{\alpha})$ and $u^2=\mathcal{O}((t_c-t)^{2\alpha})$.  Therefore, the source terms $u$ and $u^2$ in (\ref{eq:radialPR}) do not contribute to leading order, and instead the self-similar behaviour is driven by nonlinear diffusion via the Porous Medium equation
\begin{equation}
\frac{\partial u}{\partial t}= \frac{1}{r}\frac{\partial}{\partial r}\left(ru^n\frac{\partial u}{\partial r}\right),
\quad r>R(t).
\label{eq:radialPM}
\end{equation}
Furthermore, we even can keep the first source term $u$ so that
\begin{equation}
\frac{\partial u}{\partial t} \sim \frac{1}{r}\frac{\partial}{\partial r}\left(ru^n\frac{\partial u}{\partial r}\right)
+ u,
\quad r>R(t);
\label{eq:radialPRwithsource}
\end{equation}
in this case, the change of variables
\begin{equation}
\tilde{u}=\mathrm{e}^{(t_c-t)}u,
\quad
\tilde{t}=\frac{1}{n}\left(\mathrm{e}^{-n(t_c-t)}-1\right)
\label{eq:change}
\end{equation}
leads again to the Porous Medium equation
\begin{equation}
\frac{\partial \tilde{u}}{\partial \tilde{t}}= \frac{1}{r}\frac{\partial}{\partial r}\left(r\tilde{u}^n \frac{\partial \tilde{u}}{\partial r}\right),
\quad r>R(\tilde{t}).
\label{eq:radialPMtilde}
\end{equation}
We shall continue with (\ref{eq:radialPM}) but simply note that our analysis of the Porous Medium equation holds exactly for (\ref{eq:radialPRwithsource}).

\vspace{2ex}
By substituting (\ref{eq:dudt}) and (\ref{eq:d2udr2}) into (\ref{eq:radialPM}), we find that
\begin{equation}
\alpha=\frac{2\beta-1}{n},
\label{eq:alphabeta}
\end{equation}
\begin{equation}
-\left(\frac{2\beta-1}{n}\right)U+\beta\rho \frac{\mathrm{d}U}{\mathrm{d}\rho}
= \frac{1}{\rho} \frac{\mathrm{d}}{\mathrm{d}\rho}\left(\rho U^n \frac{\mathrm{d}U}{\mathrm{d}\rho}\right),
\quad \rho>\mu.
\label{eq:mainODE}
\end{equation}
Here the similarity exponent $\beta$ cannot be determined by dimensional analysis of the governing equation (\ref{eq:radialPM}) or by applying global conservation of mass; instead, it acts as an eigenvalue for a boundary-value problem associated with (\ref{eq:mainODE}), which we discuss below in some detail.  This self-similarity of the second kind makes the problem rather challenging.

\subsection{Relationship to other studies}

At this point it is worth making two comments on this formulation.  First, the hole-closing problem for the Porous Medium equation (\ref{eq:radialPM}) has been studied by a number of authors \cite{aronsongraveleau93,angenentaronson95a,aronsonetal03,beteluetal00} (including for the special case $n=3$~\cite{angenentaronson95b,diezetal98}, which corresponds to an inwardly-filling viscous gravity current).  In all of these studies, the governing equation (\ref{eq:radialPM}) is rewritten using the so-called pressure variable $v=u^n$ and the subsequent analysis involves similarity solutions of the form $v= (t_c-t)^{2\beta-1} V(\rho)$.  Further changes of variables are required to complete the analysis, which to a certain extent has the effect of hiding the physical interpretation.

Second, the hole-closing problem for the fourth-order analogue of (\ref{eq:radialPM}), the thin film equation
\begin{equation}
\frac{\partial u}{\partial t} = - \frac{1}{r}\frac{\partial}{\partial r}\left(ru^n\frac{\partial^3 u}{\partial r^3}\right),
\quad r>R(t),
\label{eq:radialthinfilm}
\end{equation}
has been studied recently by Zheng et al.~\cite{zhengetal18,zhengetal18b}.  This analogue shares some features of the hole-closing problem for (\ref{eq:radialPM}).  In the spirit of \cite{zhengetal18}, and in an attempt to provide a more transparent analysis than that which uses the pressure variable $v=u^n$, we shall concentrate on the dependent variable $u$ (and not $v$) and note some similarities and differences with \cite{zhengetal18} later.

\subsection{Far-field conditions}

To formulate the appropriate boundary-value problem associated with (\ref{eq:mainODE}), we first derive the far-field condition as $\rho\rightarrow\infty$.  As we do not want ${\partial u}/{\partial t}\rightarrow\infty$ as $t\rightarrow t_c^-$ for a fixed $r>0$ (in other words, our solution cannot blow up at any point other than $r=0$), we conclude from (\ref{eq:dudt}) and (\ref{eq:alphabeta}) that
\begin{equation}
-\left(\frac{2\beta-1}{n}\right)U+\beta\rho \frac{\mathrm{d}U}{\mathrm{d}\rho}\sim 0 \quad\mbox{as}\quad \rho\rightarrow\infty,
\end{equation}
which implies
\begin{equation}
U\sim a \rho^{(2\beta-1)/n\beta} \quad\mbox{as}\quad \rho\rightarrow\infty.
\label{eq:farfield0}
\end{equation}
This argument is used extensively in dealing with similarity solutions \cite{eggersfontelos,eggersbook}.  The constant $a$ in (\ref{eq:farfield0}) is arbitrary; however, the differential equation (\ref{eq:mainODE}) is invariant under the transformation
\begin{equation}
U\rightarrow \epsilon U, \quad \rho\rightarrow \epsilon^{n/2}\rho,
\label{eq:invariance}
\end{equation}
thus, without loss of generality, we can set $a=1$ and recover any solution we like by stretching $U$ and $\rho$ appropriately.  Therefore we have as our far-field condition,
\begin{equation}
U\sim \rho^{(2\beta-1)/n\beta} \quad\mbox{as}\quad \rho\rightarrow\infty.
\label{eq:farfield1}
\end{equation}
Numerically, we will need to truncate $0<\rho<\infty$ to $0<\rho<\rho_\infty$, and then interpret (\ref{eq:farfield1}) as two boundary conditions
\begin{equation}
U= \rho_\infty^{(2\beta-1)/n\beta}, \quad\mbox{on}\quad \rho=\rho_\infty,
\label{eq:farfield2}
\end{equation}
\begin{equation}
\frac{\mathrm{d}U}{\mathrm{d}\rho}=\left(\frac{2\beta-1}{n\beta}\right)\rho_\infty^{(2\beta-1-n\beta)/n\beta}, \quad\mbox{on}\quad \rho=\rho_\infty.
\label{eq:farfield3}
\end{equation}
Thus we can treat (\ref{eq:mainODE}) with (\ref{eq:farfield2})-(\ref{eq:farfield3}) as an initial-value problem starting at $\rho=\rho_\infty$.

\subsection{Shooting method and near-field conditions}

For a fixed $n$, we have a one-parameter family of initial-value problems, each for a different value of $\beta$.  Our strategy is to interpret these via a shooting method, where we start at $\rho=\rho_\infty$ and shoot backwards until
\begin{equation}
U=0 \quad \mbox{on}\quad \rho=\mu.
\label{eq:U0}
\end{equation}
We wish to determine the appropriate value of $\beta$ for which (\ref{eq:consmassradial}) is satisfied, or alternatively,
\begin{equation}
U^n\frac{\mathrm{d}U}{\mathrm{d}\rho}\rightarrow 0\quad \mbox{as}\quad \rho\rightarrow \mu^+.
\label{eq:dU0drho}
\end{equation}
In addition to determining the appropriate value of $\beta$, the parameter $\mu$ must also be computed as part of the numerical solution.  It will turn out that for each value of $n$, there are infinitely many pairs $(\mu,\beta)$ that satisfy (\ref{eq:farfield2})-(\ref{eq:farfield3}) and (\ref{eq:U0}), but only one pair $(\mu_c,\beta_c)$ which also satisfies (\ref{eq:dU0drho}).  In order to proceed, we need to analyse possible behaviours of solutions to (\ref{eq:mainODE}) as $U\rightarrow 0^+$.

\vspace{1ex}
\noindent
{\bf Case 1, $\beta<\beta_c$.}  By considering (\ref{eq:mainODE}) directly, we see that, provided $\mu\neq 0$, for (\ref{eq:U0}) to hold we must have
\begin{equation}
\beta \mu \frac{\mathrm{d}U}{\mathrm{d}\rho}
\sim \frac{\mathrm{d}}{\mathrm{d}\rho}\left(U^n \frac{\mathrm{d}U}{\mathrm{d}\rho}\right)
\quad \mbox{as}\quad \rho\rightarrow \mu^+.
\end{equation}
Integrating, we find
\begin{equation}
\beta \mu U \sim U^n \frac{\mathrm{d}U}{\mathrm{d}\rho}+\bar{C}
\quad \mbox{as}\quad \rho\rightarrow \mu^+,
\label{eq:barC}
\end{equation}
where $\bar{C}$ is a constant.  In the generic case where $\bar{C}\neq 0$, the two terms on the right-hand side balance, giving
\begin{equation}
U\sim C(\rho-\mu)^{1/(n+1)} \quad\mbox{as}\quad \rho\rightarrow \mu^+.
\label{eq:case1}
\end{equation}
Following~\cite{zhengetal18}, we call this case generic touch-down.

\vspace{1ex}
\noindent
{\bf Case 2, $\beta=\beta_c$.}  For the special case in (\ref{eq:barC}) in which $\bar{C}=0$, the left-hand side must balance the first term on the right-hand side, giving
\begin{equation}
U\sim (\beta_c\mu_cn)^{1/n}(\rho-\mu_c)^{1/n} \quad\mbox{as}\quad \rho\rightarrow \mu_c^+.
\label{eq:case2}
\end{equation}
Note in this case, (\ref{eq:dU0drho}) is satisfied, so this is the physically relevant solution we are after.  We call this nongeneric touch-down~\cite{zhengetal18}.

\vspace{1ex}
\noindent
{\bf Case 3, $\beta>\beta_c$.}
Here we now have $\mu=0$, which means that all the terms in (\ref{eq:mainODE}) balance in the limit.  By employing a power-law ansatz, we find that
\begin{equation}
U\sim \left(\frac{n}{4(n+1)}\right)^{1/n}\rho^{2/n} \quad\mbox{as}\quad \rho\rightarrow 0^+.
\label{eq:case3}
\end{equation}
We call this touch-down at the origin~\cite{zhengetal18}.

\vspace{1ex}
In summary, the numerical task is to vary $\beta$ until the special borderline case 2 (nongeneric touch-down) is determined for $\beta=\beta_c$ and $\mu=\mu_c$.  To ensure the borderline case is accurately identified, we utilise a high-order implicit finite-difference scheme to discretise (\ref{eq:mainODE}), with local error tolerance set to near machine precision.  The contrasting touch-down behaviours of cases 1 and 3 provide a convenient means of bracketing the critical value $\beta_c$, and hence convergence to this value is achieved iteratively through repeated bisection of the interval.

\begin{figure}[p]
\centering
\includegraphics[width=0.8\textwidth]{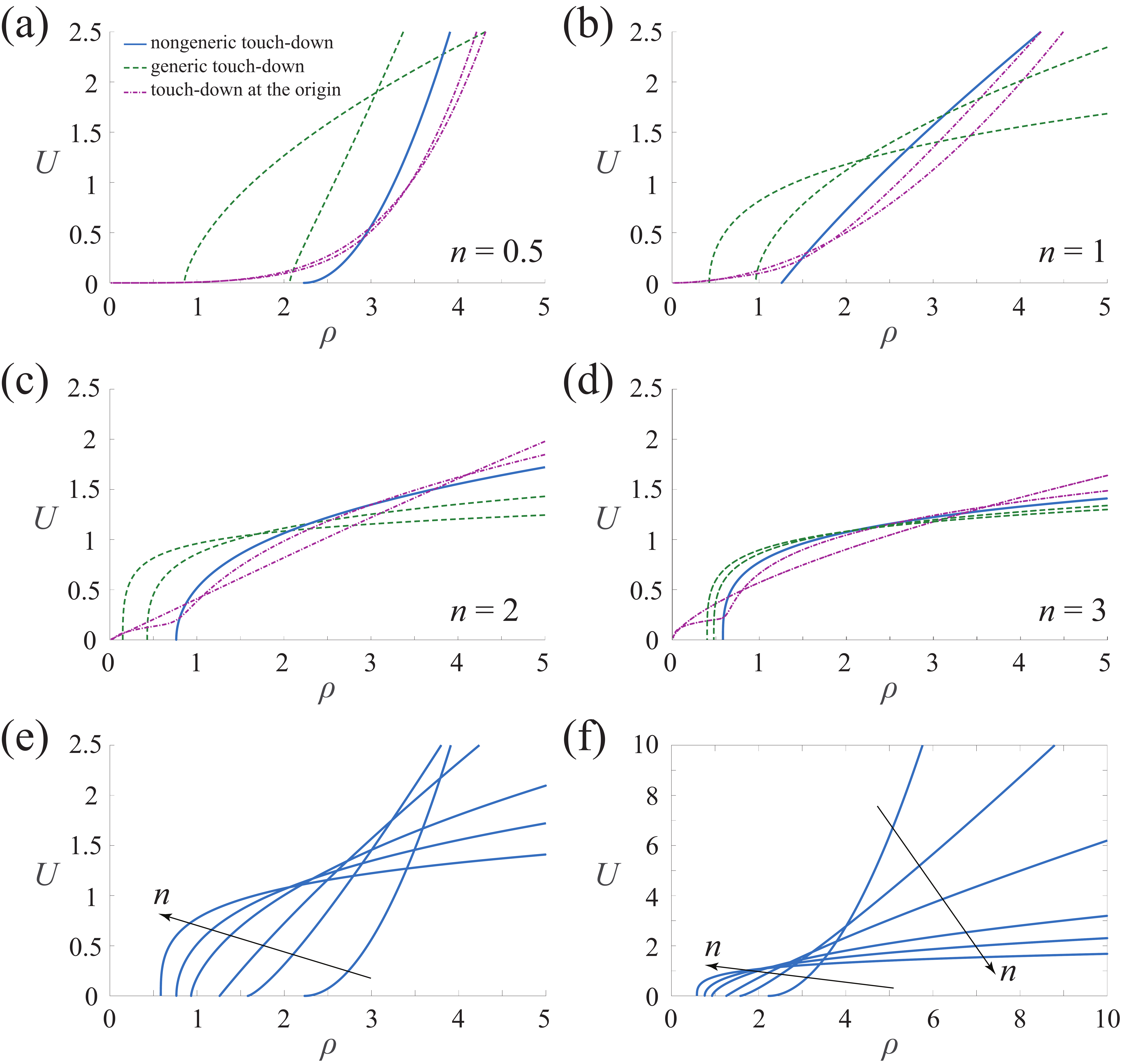}
\caption{{\bf Similarity solutions generated by the shooting method.}  (a) is for $n=0.5$.  Here the dashed (green) curves, for $\beta=0.6$ and 0.75, are examples of generic touch-down; the solid (blue) curve, for $\beta=\beta_c=0.909$, is the physically relevant nongeneric touch-down solution; while the dot-dashed (magenta) curves, for $\beta=1.05$ and 1.2, are examples of touch-down at the origin.  Using the same convention: (b) is for $n=1$ and $\beta=0.6$, 0.7, 0.856, 1 and 1.5; (c) is for $n=2$ and $\beta=0.58$, 0.65, 0.796, 0.9 and 1.5; while (d) is for $n=3$ and $\beta=0.67$, 0.7, 0.762, 0.85 and 1.5.  In (e)-(f), the physically relevant nongeneric touch-down solutions are shown for $n=0.5$ ($\beta=\beta_c=0.909$), 0.75 (0.880), 1 (0.856), 1.5 (0.822), 2 (0.796) and 3 (0.762), with the black arrow black arrows indicating increasing $n$.}
\label{fig:figure2}
\end{figure}

Some results of this task are presented in Figure~\ref{fig:figure2}.  In Figure~\ref{fig:figure2}(a) the plots are for $n=0.5$.  We can see clearly see the two representative profiles for $\beta<\beta_c$ (generic touch-down) have infinite slope at $\rho=\mu$; indeed, for this value of the diffusion exponent $n$, we have $U\sim C(\rho-\mu)^{2/3}$.  On the other hand, the two representative profiles for $\beta>\beta_c$ (touch-down at the origin) are very flat in the limit (here $U\sim \rho^4/144$).  The borderline case $\beta=\beta_c$ has $U$ scaling like $(\rho-\mu)^2$, which also has a zero slope, but does not approach the origin.  The other three examples of $n$, shown in Figure~\ref{fig:figure2}(b)-(d), also show profiles for each of the three cases described above.  We see that qualitatively the behaviour changes, depending on the value of the exponents in (\ref{eq:case1})-(\ref{eq:case3}).  In Figure~\ref{fig:figure2}(e)-(f) we show only the physically relevant similarity solutions with $\beta=\beta_c$.  Again, the qualitative behaviour as these curves intersect the $\rho$-axis depends on the exponent in (\ref{eq:case2}).  In particular, we observe that the similarity solutions for $n=0.5$ and $0.75$  have zero slope at the contact line, while the solutions for $n=1.5$, $2$ and $3$ have infinite slope.  The borderline case is $n=1$, where the slope is finite.

\begin{figure}[p]
\centering
\includegraphics[width=0.9\textwidth]{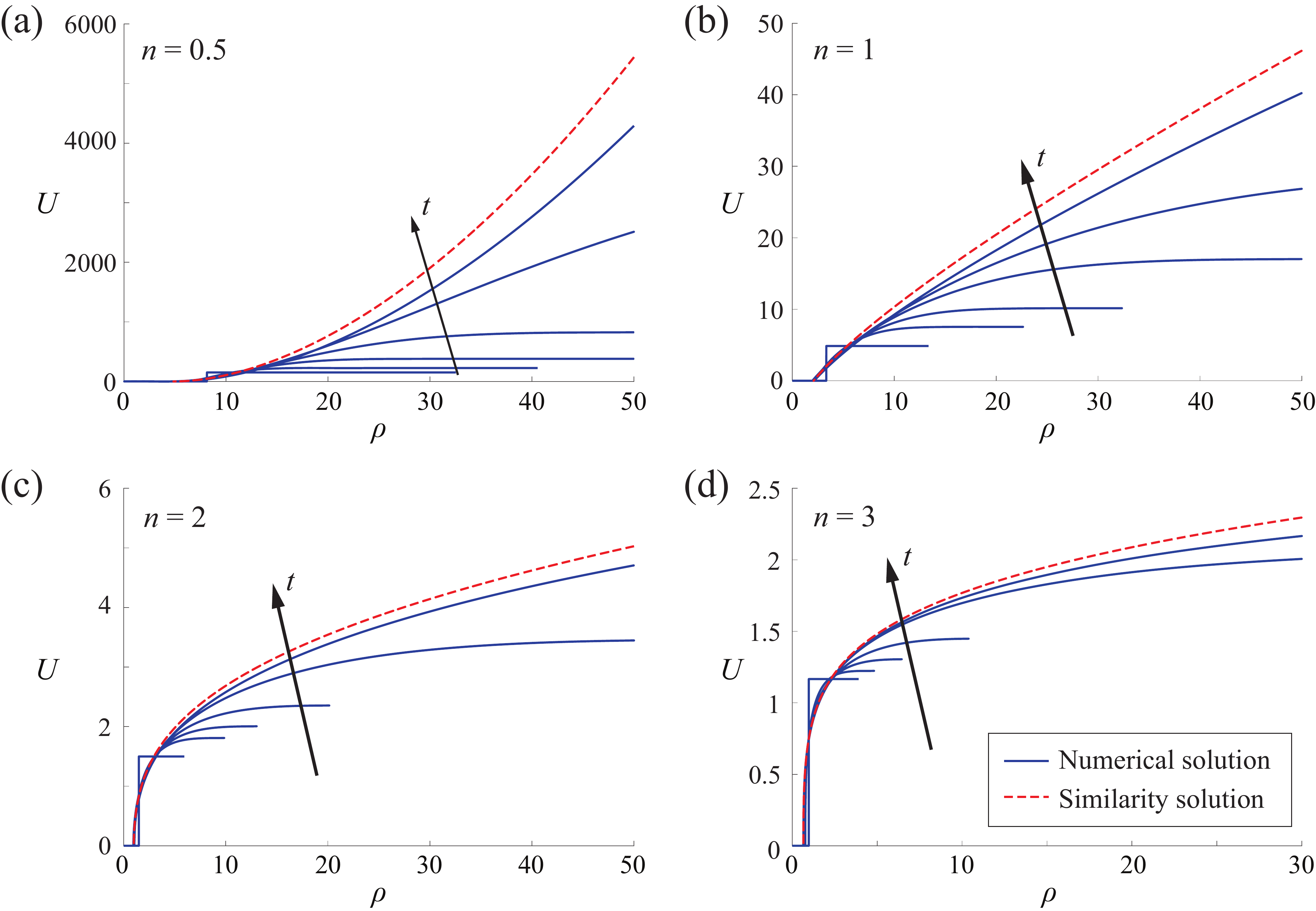}
\caption{{\bf Scaled solutions of the dimensionless Porous-Fisher equation for $n=0.5$, $1$, $2$, and $3$.}  In all cases, numerical solutions of
(\ref{eq:radialPR})-(\ref{eq:consmassradial}) with (\ref{eq:radialIC}), computed via the numerical scheme summarised in the Supplementary Material, are represented by (blue) solid curves.  Numerical solutions shown for: (a) $t = 0, 0.01, 0.02, 0.03, 0.04,$ and 0.046; (b) $t = 0.05, 0.07, 0.09, 0.10,$ and 0.108; (c) $t = 0.12, 0.16, 0.20, 0.24,$ and 0.254; and (d) $t = 0.10, 0.20, 0.30, 0.40,$ and 0.41.  In each of the four cases, solutions are obtained on $0 < r < 2$ with a nonuniform mesh of 4001 nodes, with an initial circular hole with $R(0) = 0.5$.  Zero net flux boundary conditions are imposed at both boundaries $r=0$ and $2$.  Also included in these results are the similarity solutions computed using the algorithm described in Section~\ref{sec:selfsimilar}, represented by (red) dashed curves. We see the numerical solutions approach the similarity solutions in the limit $t\rightarrow t_c^-$.
}
\label{fig:figure3}
\end{figure}

As a check on our similarity solutions, we present in Figure~\ref{fig:figure3} profiles of $U$ versus $\rho$
(where we recall that $U=u(r,t)/(t_c-t)^{(2\beta-1)/n}$ and $\rho=r/(t_c-t)^\beta$) that are computed with our numerical scheme for various times.  The figure includes plots for four values of $n$, namely $n=0.5$, $1$, $2$ and $3$.  For each value of $n$, we see the initial condition is a scaled version of the Heaviside function (\ref{eq:radialIC}), while as time increases, the numerical solutions of (\ref{eq:radialPR}) (blue solid lines) approach the similarity solution (red dashed line) in the limit $t\rightarrow t_c^-$.  This comparison between our numerical solution and our similarity solution provides confidence that our analysis in Section~\ref{sec:selfsimilar} is correct.  Note we had to rescale our similarity solutions using (\ref{eq:invariance}) in order to match the point at which $U=0$ with the numerical solution.

We close this subsection by recalling the analogue problem for the thin film equation (\ref{eq:radialthinfilm}).  As explained by Zheng et al.~\cite{zhengetal18}, the (circular) hole-closing problem for this fourth-order equation is similar to our problem (which is essentially for the Porous Medium equation (\ref{eq:radialPM})) in that there are self-similar solutions of the second kind.  Further, our procedure for computing these similarity solutions is based on that presented in \cite{zhengetal18}.  In particular, like in \cite{zhengetal18}, we use a type of shooting method to explore numerical solutions for a range of parameter values and choose the relevant solution by ensuring the near-field limiting behaviour matches a physical constraint.  The main difference between our study and that in \cite{zhengetal18} is that we have one free parameter $\beta$ and only three near-field options (which we call case 1, 2 and 3).  On the other hand, the thin film equation is higher order and so Zheng et al.'s study is more complicated.  They have a free parameter in the far-field condition as well as the similarity exponent $\beta$ and, consequently, they have more than three options for their near-field behaviour.  Another further complication for the thin film model in \cite{zhengetal18} is that there needs to be a prewetting film which regularises a well-known singularity in stress at the moving contact line.

\subsection{Similarity solutions with $k$-fold symmetry}

For the Porous Medium equation, the stability of the radially symmetric similarity solutions (\ref{eq:similaritysoln}) is known to depend on the diffusion exponent $n$.  In particular, if we consider stability of the interface $R(t)=\mu(t_c-t)^\beta$ by adding a small perturbation $\gamma(t)\cos k\theta$, then for each $k\geq 3$, the perturbed solution is stable for $n>n_k$ and unstable for $n<n_k$, where $n_k$ is some borderline exponent that can be computed numerically \cite{aronsonetal03,beteluetal00}.  By stable, we mean that $\gamma/R\rightarrow 0$ as $t\rightarrow t_c^-$.

\begin{figure}[p]
\centering
\includegraphics[width=0.8\textwidth]{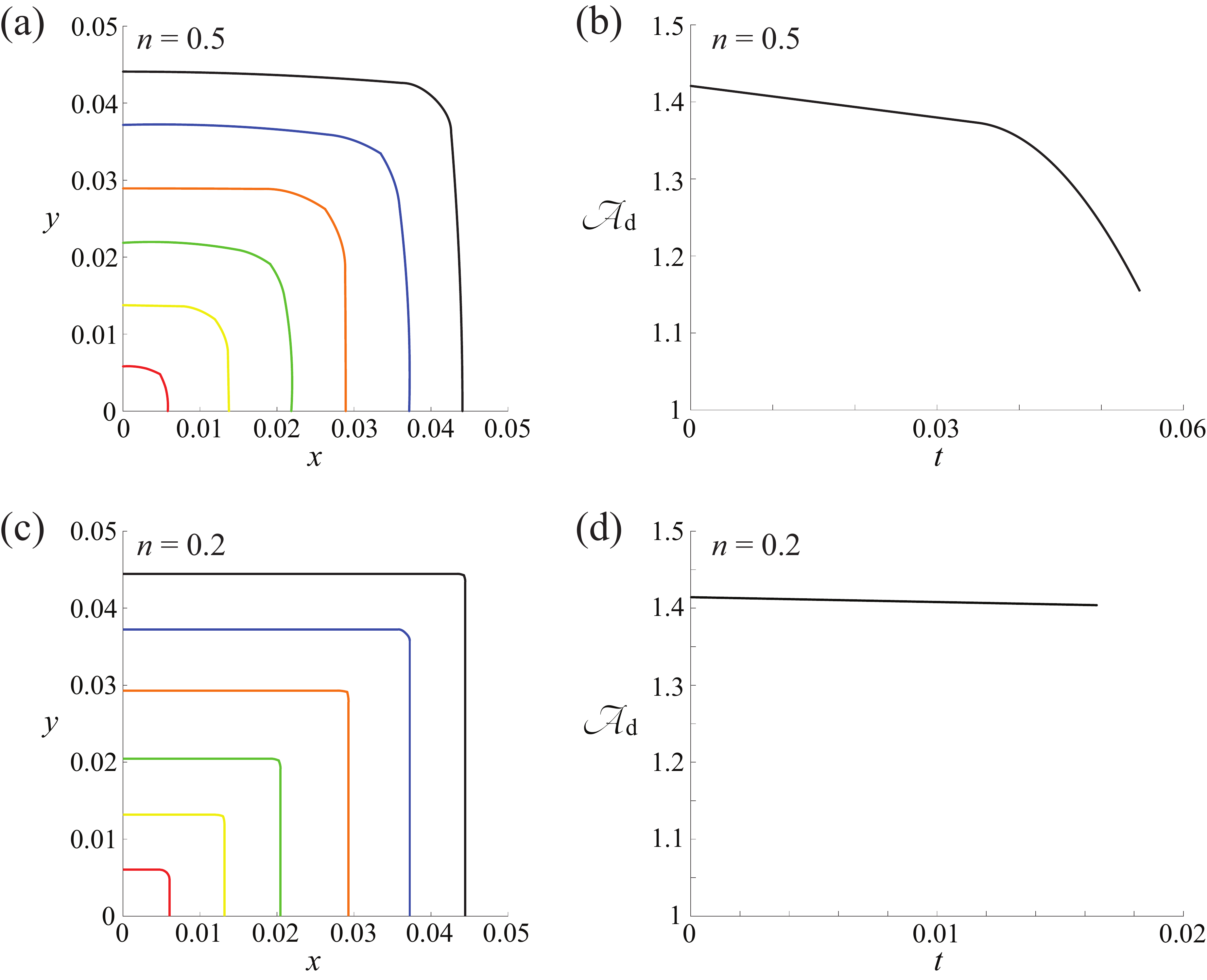}
\caption{{\bf Time evolution of the hole shape for $\mathbf{n = 0.2}$ and 0.5.} (a) Evolution of the hole shape during closing for $n = 0.2$ from $t - t_\text{c} = -5.05 \times 10^{-4}$ to $-2.5 \times 10^{-5}$. (b) Time evolution of the diagonal aspect ratio for $n = 0.2$. (c) Evolution of the hole shape during closing for $n = 0.5$ from $t - t_\text{c} = -1.09 \times 10^{-3}$ to $-9.35 \times 10^{-5}$. (d) Time evolution of the diagonal aspect ratio for $n = 0.5$. Both of the numerical solutions are obtained by solving the two-dimensional (dimensionless) Porous-Fisher equation on $0 < x < 2$ and $0 < y < 2$ with a $401\times 401$ nonuniform mesh, with an initial square hole of length 0.5 centred at origin.}
\label{fig:figure4}
\end{figure}

We postulate that the same type of stability holds for the full Porous Fisher equation~(\ref{eq:PMEdimensionless}), and now consider a relevant example for $k=4$.  Suppose the initial condition is a square-shaped hole with a side of length unity.  Here the solution has a 4-fold symmetry, so there are two options for the shape of the interface in the limit it closes.  The first is for $n>n_4$, where we assume $n_4\approx 0.32$ \cite{beteluetal00}.  For example, in Figure~\ref{fig:figure4}(a) we show the shape of the interface for a numerical solution with $n=0.5$.  Since the radially symmetric similarity solution (\ref{eq:similaritysoln}) is stable for $n=0.5$ to a 4-fold perturbation, we expect the interface for the full solution to approach a circle at extinction.  While it is difficult to compute such two-dimensional solutions accurately near extinction, our numerical results in Figure~\ref{fig:figure4}(a) appears to show the hole becoming more circular as it closes.  To support this idea, we have in Figure~\ref{fig:figure4}(b) plotted a kind of aspect ratio $\mathcal{A}_\mathrm{d}$, which is the ratio of the diagonal of the hole to the $x$-intercept.  Again, it is difficult to tell given the scales involved, but it is not unreasonable to believe that $\mathcal{A}_\mathrm{d}$ is tending to unity at extinction, which should happen if it is becoming more circular.

Also shown in Figure~\ref{fig:figure4} is an example for $n<n_4$, which demonstrates the second option for the shape of the hole as it closes.  In Figure~\ref{fig:figure4}(c)-(d) the results are for $n=0.2$, for which we postulate the interface is unstable.  Given the solution is rotationally symmetric (with 4-fold symmetry), this instability manifests itself by forcing the hole to approach a (noncircular) 4-fold symmetric shape at extinction, which is like a square with rounded corners.  This type of evolution is suggested in Figure~\ref{fig:figure4}(c).  Further, such a 4-fold symmetric shape will have a value of $\mathcal{A}_\mathrm{d}$ which is not unity.  The time-dependence of $\mathcal{A}_\mathrm{d}$ in Figure~\ref{fig:figure4}(d) appears to support the idea that $\mathcal{A}_\mathrm{d}\not\to 1$ in the hole-closing limit.

We expect that qualitatively similar results could be presented for any $k$-fold symmetric initial condition, where $k\geq 3$, with the hole approaching a circle in shape for $n>n_k$ and a $k$-fold symmetric shape (which is like a regular $k$-sided polygon with rounded corners) for $n<n_k$.  On the other hand, for $k=2$ we expect the similarity solution to unstable for all $n$, so that solutions with 2-fold symmetric initial conditions are no longer self-similar in the hole-closing limit \cite{angenentetal01,aronsonetal03}; instead, the interface becomes oval in shape with an increasingly large aspect ratio which scales like $\mathcal{A}=\mathcal{O}((t_c-t)^{-1/2})$.  We discuss this possibility further below in Section~\ref{sec:2fold} where we focus on a rectangular-shaped initial hole.

All of these stability results are analogous to those for contracting bubbles in a Hele-Shaw cell for which the viscous fluid is of a power-law type \cite{King2009,McCue2011} or for which there is a competition between surface tension and a kinetic-type boundary condition~\cite{Dallaston2013,Dallaston2016}.  For these problems, there are also linear stability results that show a circular interface may be stable or unstable, depending on a parameter value, leading to the existence of noncircular self-similar solutions (that evolve to shapes which appear like $k$-sided polygons with rounded corners).  Further, 2-fold symmetric perturbations are unstable for these Hele-Shaw problems, leading to interfaces that approach a slit in the hole-closing limit.

\section{Rectangular shaped wounds}\label{sec:2fold}

As just discussed, linear stability analysis of the radial similarity solutions shows that 2-fold symmetric perturbations grow in time \cite{angenentetal01}, which suggests that an initially rectangular hole will become longer and thinner as it closes.  We use this property to demonstrate how the diffusion exponent $n$ can be determined by fitting to aspect ratio data from a rectangular wound healing assay.

\subsection{Sticker assays}

We briefly summarise the sticker assays reported in~\cite{jinetal18}.  These two-dimensional wound healing assays were performed using (NIH 3T3) fibroblast cells.  A double-sided sticker was cut to a particular wound shape (circle, triangle or square) using a laser scribe, and then attached to a cell culture dish.  The fibroblast cells were placed in the dish and incubated overnight.  The sticker was removed to reveal the wound area of the required shape.  As the cells migrated into the vacant space, images were taken at various discrete times and subsequently analysed using ImageJ \cite{treloarsimpson13} to determine the wound area at each time point (see Supplementary Material).

In~\cite{jinetal18}, we simulated the sticker assays (with circular, triangular and square wounds), using a discrete random walk model on a hexagonal lattice incorporating crowding effects via an exclusion process~\cite{fernando10,jinetal16b,jinetal17} (whose continuum-limit description is the two-dimensional Fisher-KPP equation).  We estimated the cell proliferation rate $\lambda=0.036$ /h and the carrying capacity $K=1.4\times 10^{-3}$ cells/$\mu$m$^2$ by counting cells in sample regions of a corresponding proliferation assay and calibrating to the logistic growth model.  The random walk model was then used to estimate the diffusion coefficient $D=1200\pm 260$ $\mu$m$^2$/h which can be used for the Fisher-KPP equation.

\subsection{New experimental results}

\begin{figure}[p]
\centering
\includegraphics[width=0.95\textwidth]{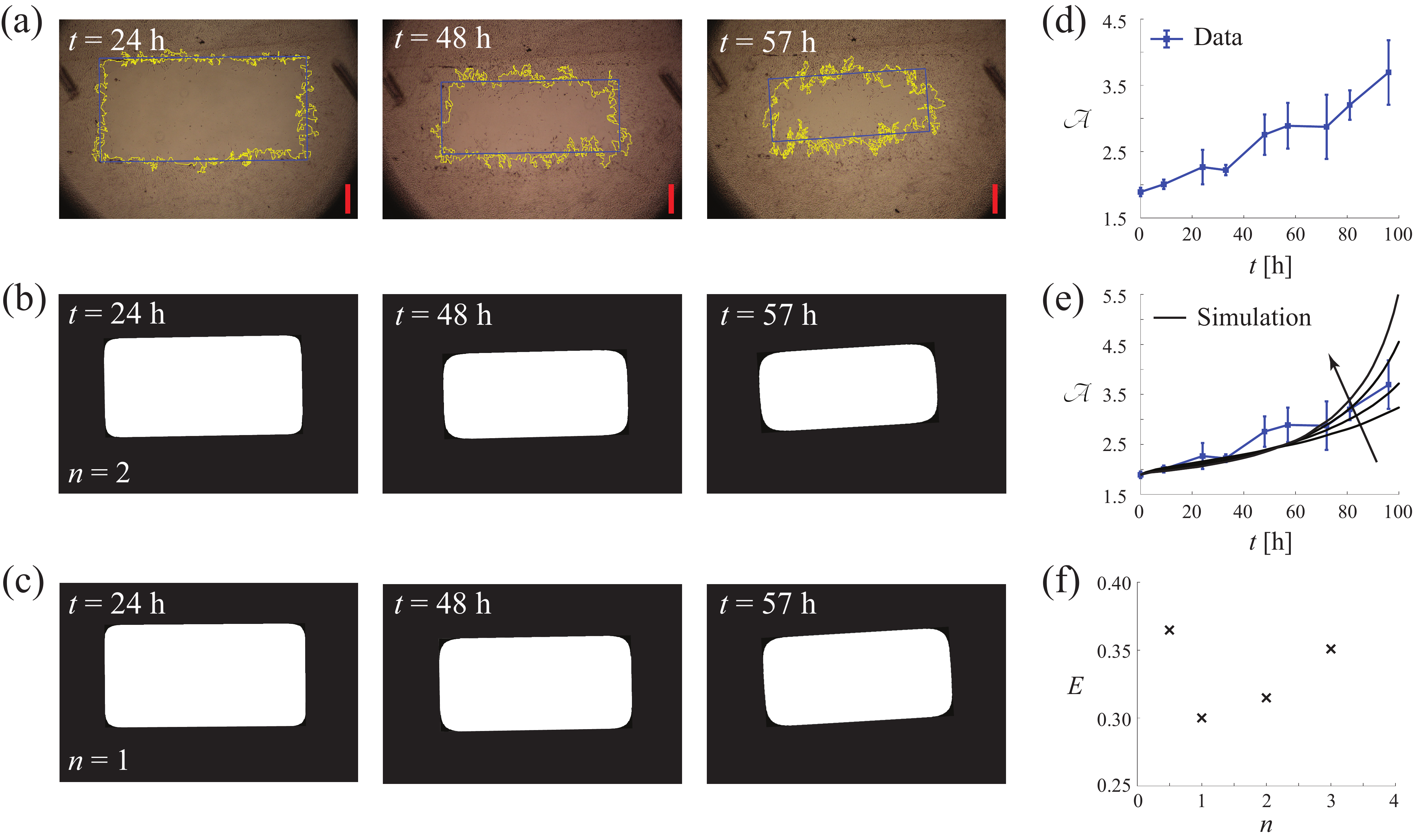}
\caption{{\bf Comparison of the aspect ratio profile between experimental data and simulation results.} (a) Experimental image of wound-healing assay at $t = 24, 48,$ and 57 h with an initial aspect ratio of 2. (b)--(c) Numerical simulation of the dimensional Porous-Fisher equation for $n = 1$ and 2, with the same initial wound size at $t = 24, 48,$ and 57 h. The white area indicates the vacant space where cell density is zero. (d) Time evolution of the aspect ratio from experimental data. (e) Comparison of the aspect ratio profile obtained by solving the Porous-Fisher equation for $n = 1, 2,$ and 3, and the experimental data. (f) The least-squares difference of the aspect ratio profile between the experimental data and simulation results for $n = 1, 2,$ and 3. Black arrow indicates the direction of increasing $n$. All the numerical solutions are obtained by solving two-dimensional Porous-Fisher equation on a domain of the same size as the experimental field of view (6 mm $\times$ 4 mm) with uniform meshes. The initial rectangular hole size is 4.20 mm $\times$ 2.22 mm. For each choice of $n$, the value of $D$ is estimated by calibrating the Porous-Fisher equation to the experimental data of circular wound \cite{jinetal18}. For all the solutions $\lambda = 0.036$ /h, $K = 1.4 \times 10^{-3}$ cells/$\mu m^2$. $\delta x = 3.75$ and $\delta y = 2.5$ for $n = 0.5, 1,$ and 2. $\delta x = 7.5$ and $\delta y = 5$ for $n = 3$.}
\label{fig:figure5}
\end{figure}

We now report on new sticker assays performed for rectangular-shaped wounds using the same protocols as in~\cite{jinetal18}.  For example, in Figure~\ref{fig:figure5}(a) we show experimental images for an initially rectangular wound whose aspect ratio is 2.  These three representative images are taken for times $t=24$, 48 and 57 h.  By approximating the wound boundary by a rectangle at each time step (see Supplementary Material), we are able to record the aspect ratio versus time for 3 replicates, and plot the result in Figure~\ref{fig:figure5}(d).

\begin{figure}[p]
\centering
\includegraphics[width=0.5\textwidth]{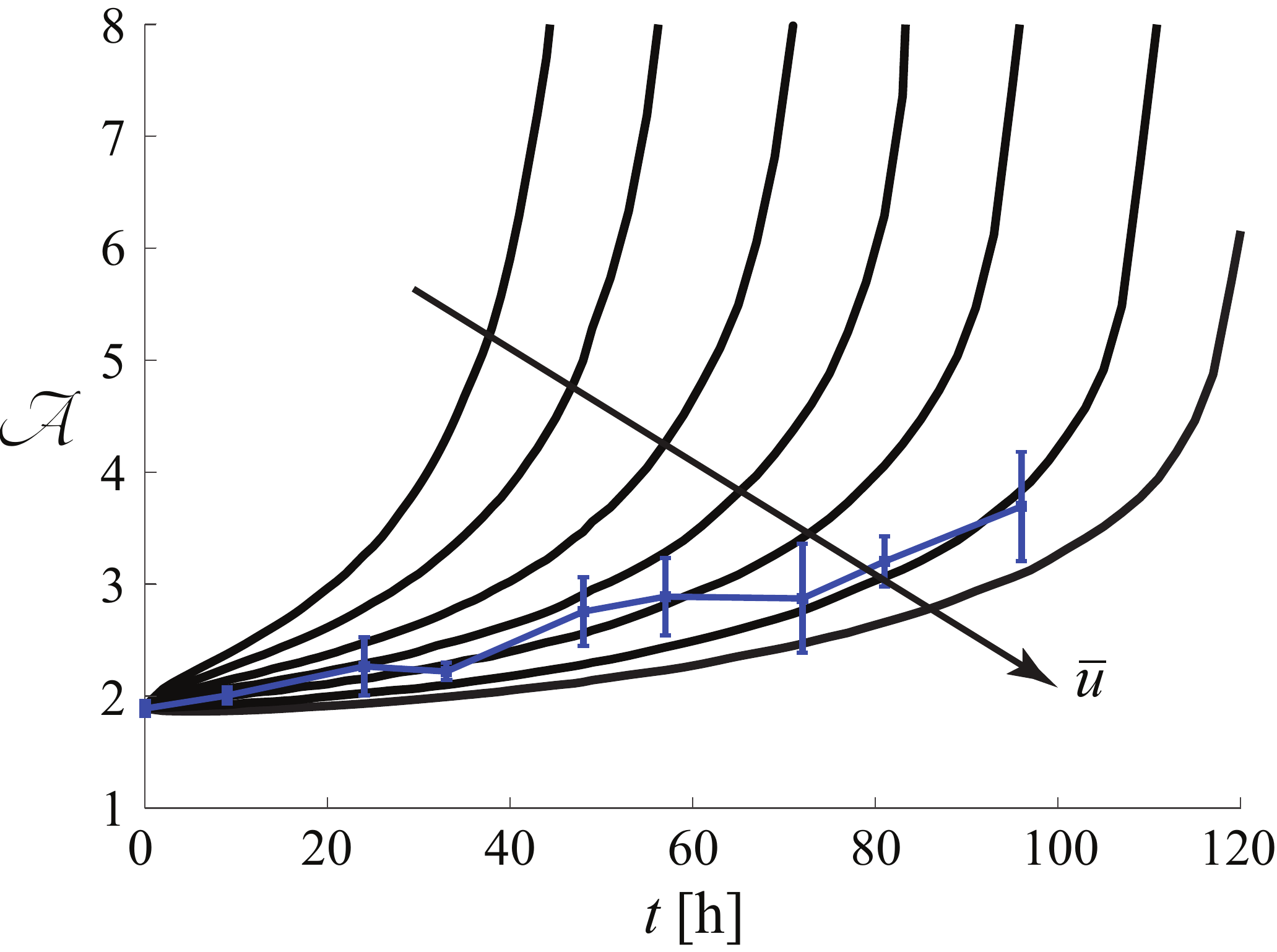}
\caption{{\bf Aspect ratios of level sets $u=\bar{u}$ for the Fisher-KPP equation.} the Fisher-KPP equation (\ref{eq:FE}) is simulated with the same initial condition as the simulation in Figure 5. Here we have used $D=1000$, $\lambda = 0.036$ /h, $K = 1.4 \times 10^{-3} $cells/$\mu m^2$. $\delta x = 7.5$, $\delta y = 5$, and $\delta t = 5 \times 10^{-3}$.}
\label{fig:figure6}
\end{figure}

\subsection{Fitting for the diffusion exponent $n$}

To calibrate our experimental data with the Porous-Fisher equation (\ref{eq:PFE}), we first take the values $\lambda=0.036$ /h and $K=1.4\times 10^{-3}$ cells/$\mu$m$^2$ from our previous study~\cite{jinetal18} described above.  Then, for each fixed value of $n$ we choose to deal with, we fit for $D$ by comparing our numerical results of (\ref{eq:radialPR})-(\ref{eq:consmassradial}) with initial condition (\ref{eq:radialIC}) with our previously obtained experimental data for the wound area with an initially circular wound~\cite{jinetal18}.  We use a simple least-squares error to identify the most appropriate choice for $D$.  Our results give estimates of $D$ and an interval of uncertainty in our estimates (Supplementary Material).

For each pair of $n$ and $D$ (together with $\lambda=0.036$ /h and $K=1.4\times 10^{-3}$ cells/$\mu$m$^2$), we then solve the Porous-Fisher equation (\ref{eq:PFE}) numerically with a rectangular-shaped hole that represents the new experimental results shown in Figure~\ref{fig:figure5}(a).  As time increases, the hole begins to close, as expected.  Representative numerical results are shown in Figure~\ref{fig:figure5}(b)-(c) for $n=1$ and $2$ (note that the initial condition for these numerical solutions involves setting $u=1$ outside of the hole, truncated for numerical purposes at finite values of $x$ and $y$).  Crucially, we plot the aspect ratio of the closing hole in Figure~\ref{fig:figure5}(e) for various values of $n$ and also include the experimental data in the same image.  This strategy allows us to choose the value of $n$ which best matches the experimental data.  A least-squares error between the experimental aspect ratio and the numerical results is shown in Figure~\ref{fig:figure5}(f).  We see that, of the values of $n$ we have simulated, the choice $n=1$ appears to provide the closest match.  Further details of the strategies employed in this subsection are provided in the Supplementary Material.

\section{Discussion}\label{sec:discussion}

In this paper we have studied various properties of solutions to the so-called hole-closing problem for the Porous-Fisher equation (\ref{eq:PFE}).  We have summarised self-similar solutions of the second kind, which apply in the neighbourhood of the moving front in the limit the hole-closes.  It turns out that the dynamics in this limit are governed by the Porous Medium equation and therefore many results carry over from previous studies of that equation~\cite{aronsongraveleau93,angenentaronson95a,aronsonetal03,beteluetal00}.  In contrast to those studies, we have formulated our analysis in terms of the original dependent variable and not the so-called pressure variable $v=u^n$; in this way our approach is more like the recent analogous study of the thin film equation~\cite{zhengetal18}.  Stability analysis \cite{angenentetal01} shows that $k$-fold (with $k\geq 3$) symmetric initial hole-shapes (like regular polygons) may become circular in the hole-closing limit, or may evolve to other non-circular shapes that keep the symmetry, all depending on the value of the exponent $n$.  Further, 2-fold symmetric initial conditions (like rectangles) will become long and thin in the limit.  Our numerical solutions confirm these predictions.

The present study is inspired by our recently published experimental data from a two-dimensional {\em sticker assay} with circular, square and triangular shaped {\em wounds} \cite{jinetal18}.  These experiments are modelled by hole-closing problems as there is an initially vacant wound area which is ultimately closed up as the cells migrate inwards and proliferate to occupy the initial wound space.  In this paper we have presented new experimental results from sticker assays with rectangular-shaped wounds.  We find the aspect ratio of the wounds increases in time in a way that agrees with our model using the Porous-Fisher equation (\ref{eq:PFE}).  For various values of $n$, we are able to fit for the diffusivity $D$ by minimising the error between the numerically computed aspect ratio and the corresponding experimental data.  Our results suggest that, for this cell line, a reasonable estimate for the exponent $n$ is $n=1$.  This result is compatible with other studies of cell migration using the Porous-Fisher equation \cite{jinetal16,warne2019}.  Furthermore, this estimate for $n$ is consistent with our previous theoretical prediction \cite{simpsonetal2011}, which suggests that for cells that themselves have an aspect ratio of $N$, the appropriate choice of $n$ is $n=N-1$.  While NIH 3T3 fibroblast cells are not at all the same shape, many have an aspect ratio of roughly two.

To put these results in context, we recall how difficult it has been to identify an appropriate choice of the diffusion exponent $n$ in (\ref{eq:PFE}) when fitting with data from wound healing assays or experiments with traditional wound shapes.  For example, for (approximately) one-dimensional fronts that arise from scratch assays with PC-3 prostate cancer cells, Jin et al.~\cite{jinetal16} calibrated the Porous-Fisher equation to the data for the examples $n=0.5$, $1$, $2$, $3$ and $4$.  While it was concluded that the choice $n=1$ outperforms the others, the evidence was not straightforward as the corresponding estimates for the diffusion coefficient $D$ varied greatly over a range of initial conditions.  In their study of circular wounds, Sherratt \& Murray~\cite{sherrattmurray90a} briefly compared numerical simulations of the Porous-Fisher (\ref{eq:PFE}) with $n=4$ with experimental results from rabbit ears~\cite{vandenbrenk1956} and attempted to fit the data for $D$.  While they observed the fit was not impressive, they did not attempt to vary $n$ but instead added other features of the model (to account for biochemical mediators).  A further example is the study of Sengers et al.~\cite{sengersetal07}, who concluded that MG-63 bone cancer cells spread out radially in a way that was well represented by the Porous-Fisher with $n=1$, but again they did not attempt to fit the data with other values of the diffusion exponent.  We conclude that our approach of using rectangular-shaped wounds in a sticker assay has the attractive feature of providing an additional means to fit for the diffusion exponent $n$ which has thus far been missing in the literature.

It is worth reflecting on how poorly the Fisher-KPP equation (\ref{eq:FE}) performs at identifying certain properties of moving fronts at they evolve.  As discussed above (and is well known), the Fisher-KPP equation does not allow solutions with compact support and so struggles to model scenarios with well-defined fronts.  In particular, we are concerned with experiments where fronts of cell populations invade an empty space, such as in our two-dimensional wound healing assays.  For this type of experiment, in order to apply the Fisher-KPP equation one must arbitrarily nominate a level set $u=\bar{u}$ as representing the moving front and then track properties of that level set as it evolves.  For example, we have solved (\ref{eq:FE}) numerically for a case with a $2\times 1$ rectangular shaped wound and plotted in Figure~\ref{fig:figure6} the aspect ratio of a number of different level sets versus time.  In all cases, the aspect ratios start at approximately $\mathcal{A}=2$ and then increase monotonically with time which is generally consistent with the experimental data (also included in this figure for reference).  However, the actual values the aspect ratios take are very different for each level set and clearly there is no obvious choice as to which level set is most appropriate.  As such, as a predictive tool, solutions to the Fisher-KPP equation (\ref{eq:FE}) are not at all useful for describing this particular experimental property.

One possible feature of our experiments that is not included in the model (\ref{eq:PFE}) is chemotaxis, whereby cells produce a chemical signal, or chemoattractant, with concentration $g$, which can promote directional motion as the cells preferentially move up or down a gradient of $g$ \cite{keller71}.  The inclusion of chemotaxis in mathematical models for wound healing and tumour growth is commonplace \cite{anderson98,olsenetal95,murphy12,menon12}.  An extension of our model (\ref{eq:PFE}) which incorporates chemotaxis could be
\begin{equation}
\frac{\partial u}{\partial t}=\,\grad\cdot\left(D\left(\frac{u}{K}\right)^n\grad u-\chi u\grad g\right)
+\lambda u\left(1-\frac{u}{K}\right),
\label{eq:chemo1}
\end{equation}
\begin{equation}
\frac{\partial g}{\partial t}=D_g\nabla^2 g+k_1 u-k_2 g,
\label{eq:chemo2}
\end{equation}
where $\chi$ is the chemotactic sensitivity coefficient, $D_g$ is the diffusivity of the chemotactic chemical, $k_1$ is the rate at which cells produce the chemotactic chemical, and $k_2$ is the rate at which the chemotactic chemical undergoes natural decay \cite{johnston2015,simpson2006}.  One significant challenge would be to obtain reasonable estimates for the four additional parameters  $D_g$, $\chi$, $k_1$ and $k_2$ by calibrating the solution of the coupled system (\ref{eq:chemo1})-(\ref{eq:chemo2}) with the experimental data.  We have not pursued this approach for two main reasons.  First, we have already been able to obtain a good match with the data using the simpler model (\ref{eq:PFE}) and we were able to use this calibrated model to draw conclusions about the use of degenerate diffusion for cell migration.  Second, as is often the case with two-dimensional wound healing assays, we have not taken any measurements of concentrations of chemoattractants; without such measurements, there is obvious metholodogy for estimating  the four additional parameters in the extended model (\ref{eq:chemo1})-(\ref{eq:chemo2}).  For this more complicated modelling to be useful, we suggest experimentalists make such measurements.

In summary, we have provided a range of evidence to support the use of nonlinear degenerate diffusion via the Porous-Fisher equation (\ref{eq:PFE}) for problems involving invading fronts.  In particular, by comparing simulations with aspect ratio data taken from sticker assays with rectangular wounds, our methodology provides a clear test for degenerate diffusion over linear diffusion in two-dimensional cell migration experiments.

\section*{Acknowledgements}
This work is supported by the Australian Research Council (DP140100249, DP170100474) and the Taiwan Ministry of Science and Technology (MOST 106-2313-B-002-031-MY3).  WJ is supported by a QUT Vice Chancellor's Research Fellowship.  SWM acknowledges many useful discussions with David Frances and Sean McElwain.  The authors are grateful for the computational resources and technical support provided by QUT's High Performance Computing and Research Support group.  Finally, the authors appreciate the supportive comments of the anonymous referees.

\end{document}